\documentclass[11pt,twocolumn,twoside]{IEEEtran}
\usepackage{amsmath}
\usepackage[margin=0.75in,headheight=0.45in]{geometry}
\usepackage[pdftex]{epsfig}
\usepackage{amsfonts}
\usepackage{amssymb}
\usepackage{fancyhdr}
\usepackage{subcaption} 
\usepackage{gensymb}
\usepackage{lipsum}
\usepackage{mathtools}
\usepackage{cuted}
\include{graphicsx}

\begin{document}
\title{\vspace{0.2in}\sc Discovering causal factors of drought in Ethiopia }
\author{Mohammad Noorbakhsh$^{1}$\thanks{Corresponding author: M. Noorbakhsh, m.noorbakhsh@warwick.ac.uk. $^1$ Mathematics Institute and Center for Complexity Science, University of Warwick, UK. $^2$ London Mathematical Laboratory, London, UK.
$^3$Institute of Mathematics and Computer Sciences, University of Sao Paulo, Brazil. }, Colm Connaughton$^{1,2}$, Francisco A. Rodrigues$^{3}$}

\maketitle
\thispagestyle{fancy}
\begin{abstract}
Drought is a costly natural hazard, many aspects of which remain poorly understood. It has many contributory factors, driving its outset, duration, and severity,  including land surface, anthropogenic activities, and, most importantly, meteorological anomalies.  Prediction plays a crucial role in drought preparedness and risk mitigation. However, this is a challenging task at socio-economically critical lead times (1-2 years), because meteorological anomalies operate at a wide range of temporal and spatial scales. Among them, past studies have shown a correlation between the Sea Surface Temperature (SST) anomaly and the amount of precipitation in various locations in Africa. In its Eastern part, the cooling phase of El Nino–Southern Oscillation (ENSO) and SST anomaly in the Indian ocean are correlated with the lack of rainfall. Given the intrinsic shortcomings of correlation coefficients, we investigate the association among SST modes of variability and the monthly fraction of grid points in Ethiopia, which are in drought conditions in terms of causality. Using the empirical extreme quantiles of precipitation distribution as a proxy for drought, We show that the level of SST second mode of variability in the prior year influences the occurrence of drought in Ethiopia. The causal link between these two variables has a negative coefficient that verifies the conclusion of past studies that rainfall deficiency in the Horn of Africa is associated with ENSO's cooling phase.
\end{abstract}
\section{Motivation}
Ethiopia, located in the east of the African continent in a region called the horn of Africa, is the most affected African country by the extreme droughts since 1900. Due to its political and socioeconomic vulnerability, it has witnessed some of the most catastrophic hydrological extreme events \cite{EMDAT}.  Since 1900, in Ethiopia, more than 77 million people were affected by drought, and more than 400,000 people died from the famine afterward \cite{EMDAT}. For this reason, accurate prediction plays a vital role in drought preparedness and risk mitigation. Thus, obtaining the influencing factors of drought, which leads to the improvement of drought forecasting models, is very important.

In general, several factors contribute to instigating rainfall deficiency in the African continent; however, the major influencing factors are ENSO and Sea Surface Temperature (SST) \cite{Masih2014APerspective}. Different phases of ENSO are associated with the occurrence of drought across various regions of Africa. While in the southern areas, droughts occur in the warm phase of ENSO (El Niño) \cite{Nicholson1997TheRainfall}, in the eastern part, it happens when the ENSO is in its cold phase (La Niña) \cite{Masih2014APerspective}. Dutra et al. (2013) \cite{Dutra2013TheProducts} found that the precipitation deficit in the Horn of Africa in October–December 2010 was associated with a robust La Niña event. Lott et al. (2013) \cite{Lott2013CanChange} also indicated that ENSO was firmly in the La Niña phase during the 2010-2011 drought. Tierney et al. (2013) \cite{Tierney2013MultidecadalOcean} argued that droughts in the Horn of Africa from 1999 to 2011 were partly due to the concurrent La Niña phase of ENSO.

Moreover, some studies investigated the correlation between drought in the Eastern part of Africa and SST in the Indian Ocean. Tierney et al. (2013) \cite{Tierney2013MultidecadalOcean}  and Funk et al. (2008) \cite{Funk2008WarmingDevelopment} suggested that the Indian Ocean influences rainfall in the Eastern part of Africa by modifying the local Walker circulation. Hasternath et al. (2007) \cite{Hastenrath2007DiagnosingAfrica} argued that rainfall deficiency in eastern Africa in 2005 happened concurrently with anomalously cold waters in the northwestern and warm anomalies in the southeastern extremity of the equatorial Indian Ocean basin. Funk et al. (2008) \cite{Funk2008WarmingDevelopment} argued that the decline in rainfall over eastern Africa is correlated with the warming of the central Indian Ocean. 

The above studies use the correlation coefficient to measure associations between drought and its potential influencing factors. The problem is correlation does not imply causation \cite{reichenbach1991direction}. Specifically, when trying to infer causality in time series data, the correlation coefficient cannot detect false causal links originated from autocorrelation within each time series and their common drivers \cite{Runge2017CausalGeosciences}. This study aims to uncover the role of variables having a causal effect with a time lag of a maximum of 12 months on drought in Ethiopia by quantifying the causal inter-dependencies of the underlying system. Besides, we seek to verify their roles in complementing predictive machine learning models \cite{Runge2019InferringSciences}, by investigating whether their presence as predictors in a forecasting model will improve the overall performance of the base model.

\section{Method}
\subsection{Data}
We use monthly precipitation data from GPCC Global Precipitation Climatology Centre \cite{schneider2011gpcc}, which covers period from Jan-1946 to Dec-2015 for Ethiopia with the resolution of 0.5\degree  in both latitude and longitude. For SST, we use monthly data from NOAA Extended Reconstructed Sea Surface Temperature (SST) V5 \cite{huang2017noaa}, from Jan-1946 to DEC-2015 with a spatial resolution of 2.0\degree in both latitude and longitude and spatial coverage of 88.0N - 88.0S, 0.0E - 358.0E.

\subsection{Defining Drought}\label{ssec:drought}
We represent drought by a univariate monthly time series, which indicates the number of grid points in drought conditions each month in Ethiopia. In this section, We follow the method introduced by McKee et al. (1993) \cite{mckee1993rel} to convert cumulative rainfall into the Standard Precipitation Index (SPI). 

Location $i$ is in drought condition in the month $m_t$ when the rain in this location from $m_{t-11}$ to $m_t$ is deficient compared to the rainfall of a reference period in the same place.The reference period for $m_t$ is the distribution of 12-months cumulative rainfall in period of 360 months before $m_{t-11}$. Since the distribution of the cumulative rainfall is not normal \cite{mckee1993rel}, we fit a Gamma probability distribution by applying maximum likelihood estimation using the SciPy library in Python \cite{2020SciPy-NMeth} to the 360-months reference period and obtain the scale and shape parameters of Gamma distribution. We then use these parameters to generate the cumulative probability for cumulative rainfall:

\begin{equation}\label{Gamma}
G(x)=\frac{\int_0^x u^{\alpha - 1}e^{\frac{-u}{\beta}}\mathrm{d}u }{\beta^{\alpha} \Gamma(\alpha)}
%\label{eq:G}
\end{equation}
where $x$ is the value of cumulative rainfall from $m_{t-11}$ to $m_t$, $\alpha$ is the scale parameter, $\beta$ is the shape parameter and $\Gamma$ is the gamma function.

Since the gamma function is undefined for zero values, and we likely encounter a zero value for the cumulative rainfall over 12 months, we convert the cumulative probability to:

\begin{equation}
      H(x) = q + (1-q)G(x)
    \label{eq:H}
\end{equation}

%\begin{equation}
%      H(x) =
%    \begin{cases}
%      q + (1-q)G(x), & \text{$x > 0$}\\
%      q, &  \text{x = 0}\\
%    \end{cases}
%   \label{eq:H}
%\end{equation}

\begin{figure}
\begin{center}
\epsfxsize=1\hsize \epsfbox{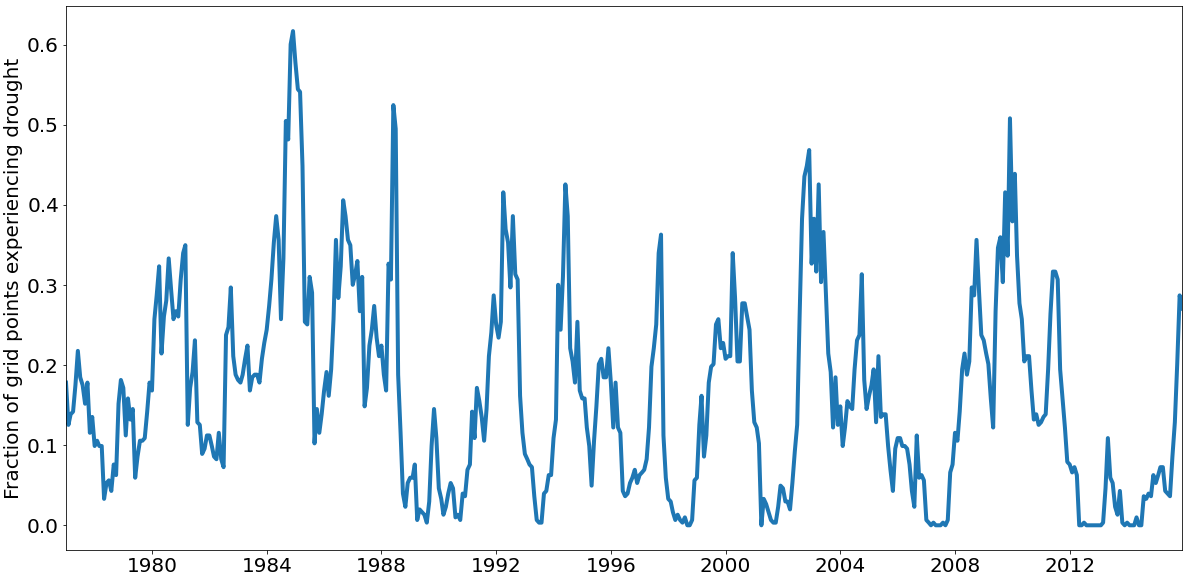}
\end{center}
\caption{Monthly fraction of grid points in drought condition in Ethiopia}
\label{Drought}
\end{figure}

Where $q$ is the probability of zero values computed from the reference period.
We then convert the cumulative probability $H(x)$ to the standard normal random variable by applying equiprobability transformation described in \cite{panofsky1958some}. The obtained value is $SPI^i_t$ for the month $t$ in the location $i$, and according to McKee et al. \cite{mckee1993rel}, when SPI reaches a value of -1 or less, the location $i$ is in drought condition in month $t$. By computing $SPI^i_t$ for $i = 1 \dots 303$, and $t = 1 \dots 468$ and counting the number of grid points with SPI equal or below -1 at each month, we obtain the final time series of drought in Ethiopia named it as drought time series in the subsequent sections. Figure \ref{Drought} displays the drought time series normalized by the number of grid points in Ethiopia.

\subsection{Detecting non-random mode of variability}
Since the global SST dataset contains 10998 time series, the computation of finding causal links among them and their lagged copies is highly expensive, and applying a dimensionality reduction techniques is necessary. We use rotated Principal Component Analysis (PCA) \cite{kaiser1958varimax}, which is the most extensively used method to solve the problem of high-dimensionality of climate datasets. It converts the original dataset into a new one with a much smaller number of dimensions and preserves most of the original data's information \cite{Hannachi2007EmpiricalReview}.  

\begin{figure}
\begin{center}
\epsfxsize=1\hsize \epsfbox{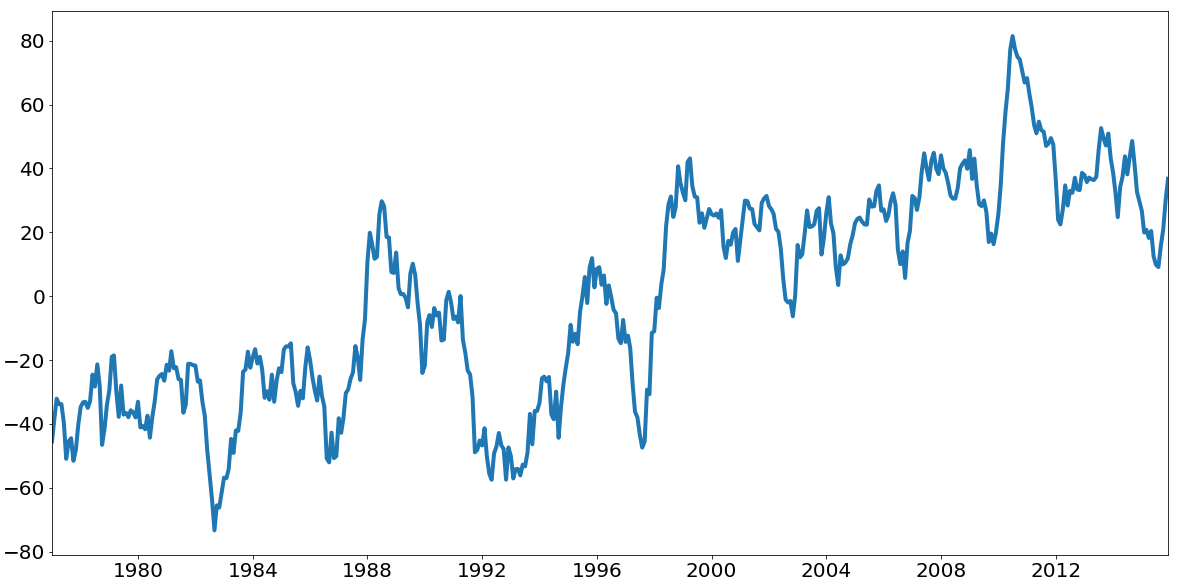}
\end{center}
\caption{The time display of the first mode of variability of SST (PC 1).}
\label{fig:PC1}
\end{figure}

\begin{figure}
\begin{center}
\epsfxsize=1\hsize \epsfbox{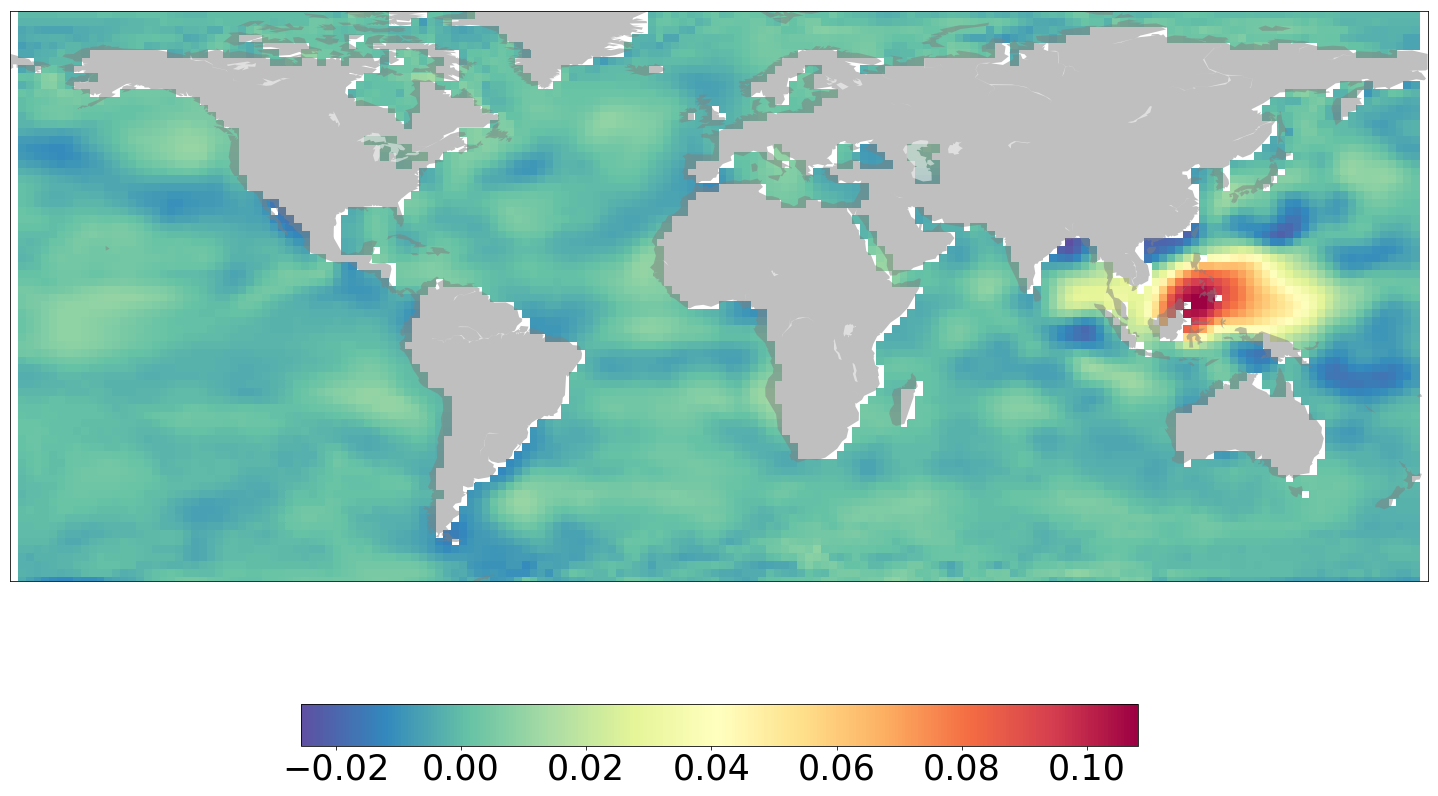}
\end{center}
\caption{The spatial structure of the first mode of variability of SST (EOF 1).}
\label{fig:EOF1}
\end{figure}

This method decomposes the data into temporal and spatial components \cite{Hannachi2007EmpiricalReview}. The temporal component is a set of uncorrelated orthogonal time series called Principal Components (PC), and the spatial component is a set of orthogonal patterns called Empirical Orthogonal Functions (EOF). Each PC has a corresponding EOF, and they are usually in descending order of explained variance in the original dataset. Figures \ref{fig:PC1} and \ref{fig:EOF1} show the time dependence (PC) and the space dependence (EOF) of SST's first mode of variability, respectively.

Rotated PCA detects modes of atmospheric variability such that PCs describe variation in time and EOFs describe the corresponding spatial patterns \cite{Huth2006ThePressure, Hannachi2007EmpiricalReview}. The number of modes is equal to the number of dimensions in the original dataset, but we are looking for "non-random modes." A mode is non-random when it is unlikely that it is generated by a set of independent, identically distributed, mean zero stochastic processes residing at grid points. Therefore such modes are likely results of correlation structures in the analyzed data \cite{Vejmelka2015Non-randomData}.

\subsubsection{Data Prepossessing}
The SST dataset is a gridded dataset composed of a space-time field, $X(t,\textbf{s})$ representing the value of SST at time t and location \textbf{s}. The dataset can be written in matrix form:

\begin{equation}
\label{eq:matrix}
X = 
 \begin{pmatrix}
  x_{1,1} & x_{1,2} & \cdots & x_{1,n} \\
  x_{2,1} & x_{2,2} & \cdots & x_{2,n} \\
  \vdots  & \vdots  & \ddots & \vdots  \\
  x_{t,1} & x_{t,2} & \cdots & x_{t,n} 
 \end{pmatrix}  
\end{equation}

We convert time series of the dataset or column vectors of the matrix in equation \ref{eq:matrix} into mean anomaly by applying phase averaging, which treats the annual cycle as a fixed mean and fixed variance effect. We divide each time series into twelve groups based on each observation's calendar month, and for each group, we subtract its mean from its members, Then we construct a single time series based on the order of the original data \cite{Donges2009ComplexMethods}. It is defined as:
\begin{equation}
\begin{aligned}
    x_i(t) \Rightarrow x_i(m,y) \\
    a_i(y,m) = x_i(m,y) - \langle x_i(m,y) \rangle_m
    \end{aligned}
\end{equation}

Where $m \in \{1,\dots,12\}$ and $\langle x_i(m,y) \rangle_m$ denotes the expected value of time series with respect to the value of the calendar month $m$. We then scale each anomaly time series by the square root of the cosine of its latitude to address the problem of unequal-area grid \cite{chung1999weighting}.

\subsubsection{Computation of EOFs and PCs}
To compute EOFs and PCs of the SST dataset, we need to compute the covariance matrix from the anomaly matrix $A$. The covariance matrix is defined by:

\begin{equation}
\label{eq:cov}
S = \frac{1}{t} A^TA
\end{equation}

%\begin{equation}
%\label{eq:cov}
%S = \frac{1}{n} \sum_{i=1}^{n}A^TA
%\end{equation}

The purpose of PCA is to find a unit-length vector $\textbf{U} =(u_1, u_2, \dots, u_n)^T$ such that $A\textbf{U}$ explains maximum variance of the original dataset. The EOFs are the solution to the following problem, which are are the eigenvectors of $S$:
\begin{equation}
\label{eq:cov}
S\textbf{U} = \lambda^2\textbf{U} 
\end{equation}

The $k^{th}$ EOF is the $k^{th}$ eigenvector $\textbf{U}_k =(u_{k1}, u_{k2}, \dots, u_{kt})^T$, and the projection of the anomaly matrix $A$ onto the $k^{th}$ EOF is the $k^{th}$ PC, $\textbf{d}_k = A\textbf{u}_k$, with variance equal to the $k^{th}$ eigenvalue, $\lambda^{2}_{k}$ of $S$ \cite{Hannachi2007EmpiricalReview}.We obtain the order of PCs and EOFs by sorting their corresponding eigenvalues in decreasing order. 

\begin{figure}
\begin{center}
\epsfxsize=1\hsize \epsfbox{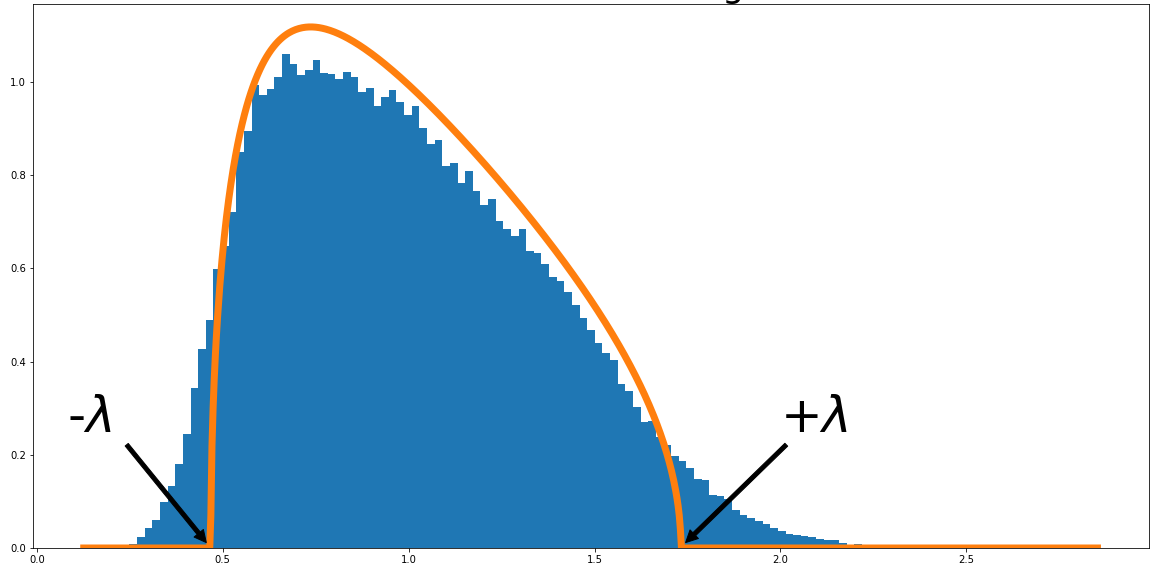}
\end{center}
\caption{Marchenko–Pastur distribution and Eigenvalues boundaries.}
\label{Marchenko–Pastur}
\end{figure}

The number of PCs and EOFs of the SST dataset is equal to the number of its time series, 10988. We use the Marchenko–Pastur distribution in the theory of random matrices to find the boundaries for eigenvalues of a correlation matrix, above which the eigenvalues are non-random \cite{marvcenko1967distribution}. Figure \ref{Marchenko–Pastur} shows the Marchenko–Pastur distribution and the boundaries for eigenvalues. The boundaries for $S$ from equation \ref{eq:cov} are defined as:

\begin{equation}
\label{eq:eigen}
\lambda_{\pm} = \left(1 \pm \sqrt{\frac{n}{t}}\right)^2
\end{equation}

By applying the above methods, 76 out of 10988 eigenvalues are non-random, and we choose the first 76 PCs and EOFs for our further analysis. 

\subsubsection{VARIMAX}
The main problem with non-rotated PCA is that the obtained EOF patterns do not have a simple structure and are not easy to interpret. This problem arises because of the strong constraints of EOFs, namely orthogonality on the EOFs/PCs. Rotating versions of PCA is an attempt to make EOF patterns more understandable by easing the constraints of orthogonality and uncorrelation \cite{Hannachi2007EmpiricalReview}.

Given a $n \times m$ EOF matrix $U_m =(\textbf{u}_{1}, \textbf{u}_{2}, \dots, \textbf{u}_{m})$, which contains $m$ leading EOFs and $m$ is 76 in our analysis, we seek to find $m \times m$ rotation matrix $R$ to obtain $B$ that:
\begin{equation}
\label{eq:REOF}
\textbf{B} = U_m\textbf{R}
\end{equation}

The most used rotation algorithm is VARIMAX, introduced by Kaiser (1958) \cite{kaiser1958varimax}. It simplifies the structure of the patterns by pushing the EOFs towards zero, or $\pm 1$ \cite{Hannachi2007EmpiricalReview}. It maximizes a simplicity criterion based on:

\begin{equation}
\label{eq:REOF}
max \left( f(B) = \sum_{k=1}^{m} \left[ p\sum_{j=1}^{p}b^4_{jk} - \left(\sum_{j=1}^{p}b^2_{jk} \right)^2 \right] \right)
\end{equation}

\begin{figure}
\begin{center}
\epsfxsize=1\hsize \epsfbox{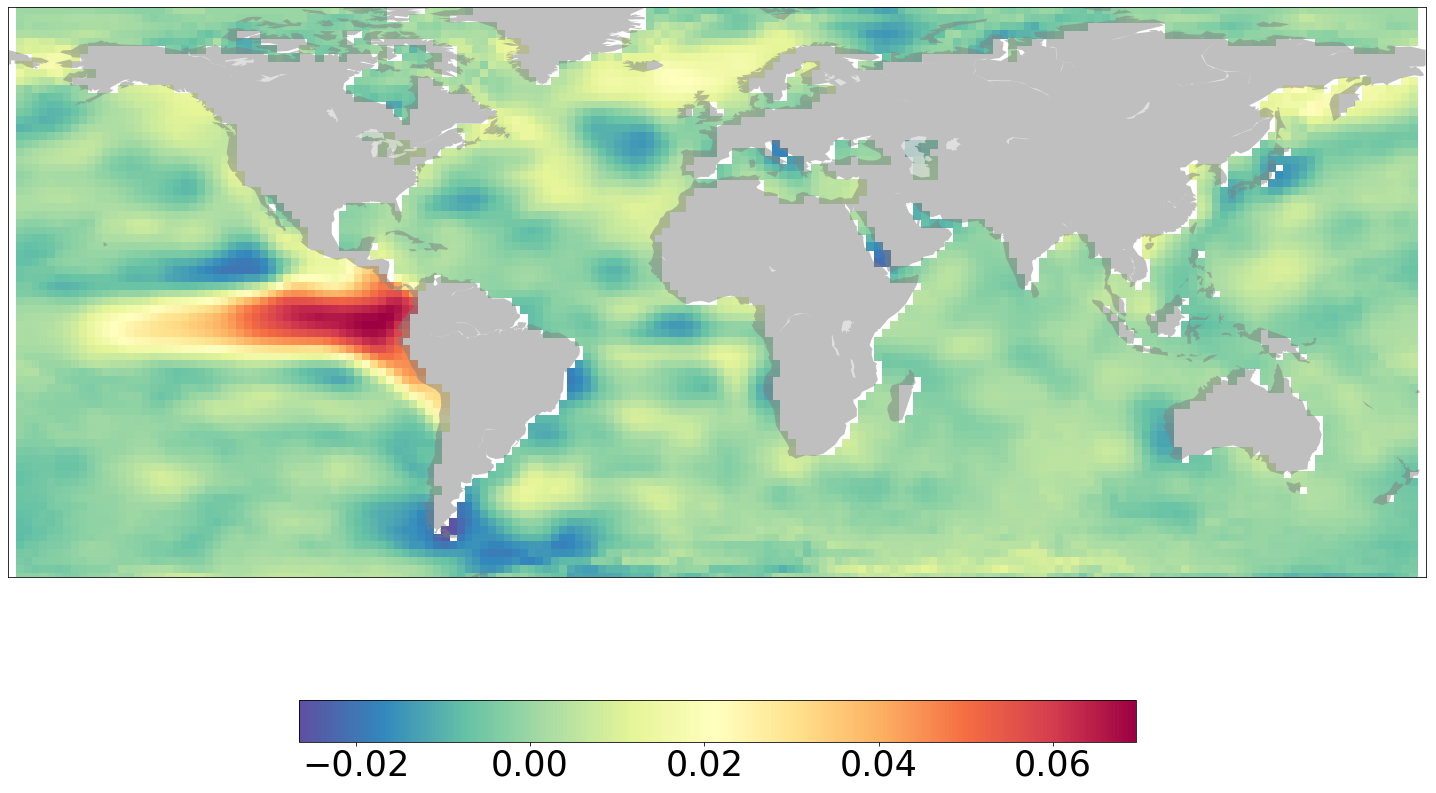}
\end{center}
\caption{The spatial structure of the second mode of variability of SST (EOF 2).}
\label{fig:EOF2}
\end{figure}

\subsection{Detecting causal relationships from time series}
The purpose of causal discovery from a finite time series sample is to remove spurious links caused by common drivers, indirect associations, and autocorrelation effects within each time series (spurious link detection) while estimating the real causal associations (high detection power).The problem is that to detect more spurious links, the model should have more variables, but this may lower detection power. An acceptable causal discovery method should maintain a trade-off between false positive rate and detection power \cite{runge2019detecting}. 

Although causation generates a statistical dependency between causes and effects, a dependency between two variables is a necessary but not sufficient condition for causality \cite{bontempi2015dependency}. The idea behind most of the causal inference frameworks is to find potential common drivers of dependent variables. If there are no drivers from the observed variables, based on the assumption of Causal Sufficiency (that there are no unobserved variables \cite{Runge2018CausalEstimation}), we can conclude that the statistical dependency is causal. The typical approach for investigating if $Z$ is the common driver of $X$ and $Y$ is to apply conditional independence: 
\begin{equation}
  X \perp \! \!\! \perp Y|Z  \Longleftrightarrow p(x,y|z) = p(x|z)p(y|z) \;  \forall x,y,z
  \label{eq:dependence}
\end{equation}

There are two extreme approaches to discover common drivers. The first one is when we do not condition on any variables (Z set is empty in equation \ref{eq:dependence}). Then this approach is similar to measures without conditioning such as Correlation Coefficient or Mutual Information(MI): 
\begin{equation}
 I(X;Y) = \int_{y} \int_{x} 
                 p(x,y) \log{ \left(\frac{p(x,y)}{p(x)\,p(y)}
                              \right)}\mathrm{d}x\mathrm{d}y   
\label{eq:MI} 
\end{equation}

These measures have a very high false-positive rate, and they cannot detect spurious links. 

The second extreme approach is when Z in equation \ref{eq:dependence} represents all other variables and their past values, including Y until $\tau_{max}$. This is called \textit{Full Conditional Independence} (FullCI) \cite{Runge2018CausalEstimation}:
\begin{equation}
  I^{FullCI}_{i \rightarrow j}(\tau) = I(X^i_{t-\tau};X^j_{t}|\mathbf{X}^{(t-1,\dots,t-\tau_{max})}_t \backslash \{X^i_{t-\tau}\})
  \label{eq:FullCI}
\end{equation}
Where $\mathbf{X}^{(t-1,\dots,t-\tau_{max})}=(\mathbf{X}_{t-1},...,\mathbf{X}_{t-\tau_{max}})$ and $I$ can be any conditional dependence measure such as Conditional Mutual Information(CMI) \cite{Runge2018CausalEstimation}:
\begin{small}
\begin{equation}
 I(X;Y|Z)=\iiint p(x,y,z) \log{\left(\frac{p(x,y|z)}{p(x|z)\,p(y|z)}
                              \right)}dxdydz   
\label{eq:CMI} 
\end{equation}
\end{small}
The second extreme approach suffers from the problem of low detection power due to "the curse of dimensionality" \cite{Runge2018CausalEstimation}.

This study uses PCMCI, a data-driven causal inference framework from time series based on the graphical causal models \cite{spirtes2000causation}. It enables researchers to discover causality networks from high-dimensional, non-linear, and autocorrelated datasets such as climate data \cite{runge2019detecting}. It takes a moderate approach to the condition set ($Z$ in equation \ref{eq:CMI}) by applying a two-step technique. The first step is called the condition-selection step that tries to reduce the number of members set to improve the problem of low detection power by finding the optimal set of variables. It uses the Causal Markov Property \cite{spirtes2000causation} from Causal Discovery Theory \cite{pearl2000models}, which indicates the parents of a variable are sufficient condition set. So in equation \ref{eq:FullCI}, both parents of $X^i_{t-\tau}$ and $X^j_{t}$ are enough to be conditioned on, and the other variables are redundant.  

In the first step, parents of all included time series variables, and their lagged copies are identified by applying $PC_1$ algorithm \cite{spirtes1991algorithm}. The second step is the momentary conditional independence (MCI):
\begin{equation}
MCI:X^i_{t-\tau}\perp \! \!\! \perp X^j_{t}|\hat{P}(X^j_{t})\backslash \{X^i_{t-\tau}\},\hat{P}(X^i_{t-\tau})
\end{equation}
Where $\hat{P}(X^j_{t})$ is representing parents of $X^j_{t}$. 

It investigates whether $X^i_{t-\tau}$ causes $X^j_{t}$ by performing independence test conditions on both their parents excluding $X^i_{t-\tau}$ if it is among the parent set of $X^j_{t}$.

By choosing the condition-selection significance level $\alpha_{pc}$, PCMCI can move toward the two extreme approaches. If $\alpha_{pc}$ is equal to zero, then the condition sets will be null, and PCMCI will be the same as Mutual Information. On the other hand, $\alpha_{pc}$ equal to one makes PCMCI a Full Conditional Independence approach. This flexibility in choosing the level of $\alpha_{pc}$ is an excellent advantage of PCMCI over other frameworks by keeping a balance between detection power and false positive rate \cite{Runge2017DetectingDatasets}.

\subsection{Assumption of Causal Discovery from Observational Time Series}
We consider results obtained from PCMCI as causal under three assumptions of \textbf{Causal Sufficiency}, \textbf{Causal Markov Condition} and \textbf{Faithfulness} \cite{runge2019detecting}. Causal Sufficiency assures that all the common causes are among the measured variables. Causal Markov Condition assumes that a variable's parents are sufficient condition set, and that variable is independent of any other variable. Faithfulness assumes that the causal graphical structure contains all observed conditional independencies.  

\section{Evaluation}
\subsection{Results}
We exploit the PCMCI algorithm's implementation in Python, named "TIGRAMITE" \cite{Runge2017DetectingDatasets}, to discover the causal links among 76 PCs of global SST data, drought time series in Ethiopia, and their lagged copy with a maximum lag of 12 months. PCMCI generates a causal time-series graph, in which we focus on the links originating from other factors to the drought time series.

We use the partial correlation for the conditional independence test. For the value of $\alpha_{pc}$, we use the built-in optimization method in TIGRAMITE, which uses the Akaike information criterion (AIC) \cite{sakamoto1986akaike} to find the best value over a range of values. The significant level, $\alpha$, for the conditional independence test in the MCI step is 0.05. 

The PCMCI algorithm identifies 52 links from other variables to the drought variable. We show the five strongest links in table \ref{table:results}. The first and the third variables are auto-dependency links, which are not causal associations, and they originate from the auto-correlation characteristic of the SPI time series. 

The strongest causal association is originating from the second PC of SST with a lag of 12 months. Figure \ref{fig:EOF2} shows the second EOF of SST that its most substantial contribution to the second PC is coming from the ENSO region in the Pacific. The negative sign of the coefficient indicates the reverse association, which implies that when the second PC of SST is low, there is a high chance that drought occurs in Ethiopia within the next year. This result verifies prior conclusions \cite{Dutra2013TheProducts, Lott2013CanChange, Tierney2013MultidecadalOcean} that rainfall deficiencies in the Horn of Africa between 1999 to 2011 and the cooling phase of ENSO or La Niña event are correlated.  We can conclude from this particular result that the association between La Niña and drought in the Horn of Africa is in the form of causality, which conveys a powerful La Niña in any year can stimulate drought conditions in Ethiopia the year after.

\begin{figure}
\begin{center}
\epsfxsize=1\hsize \epsfbox{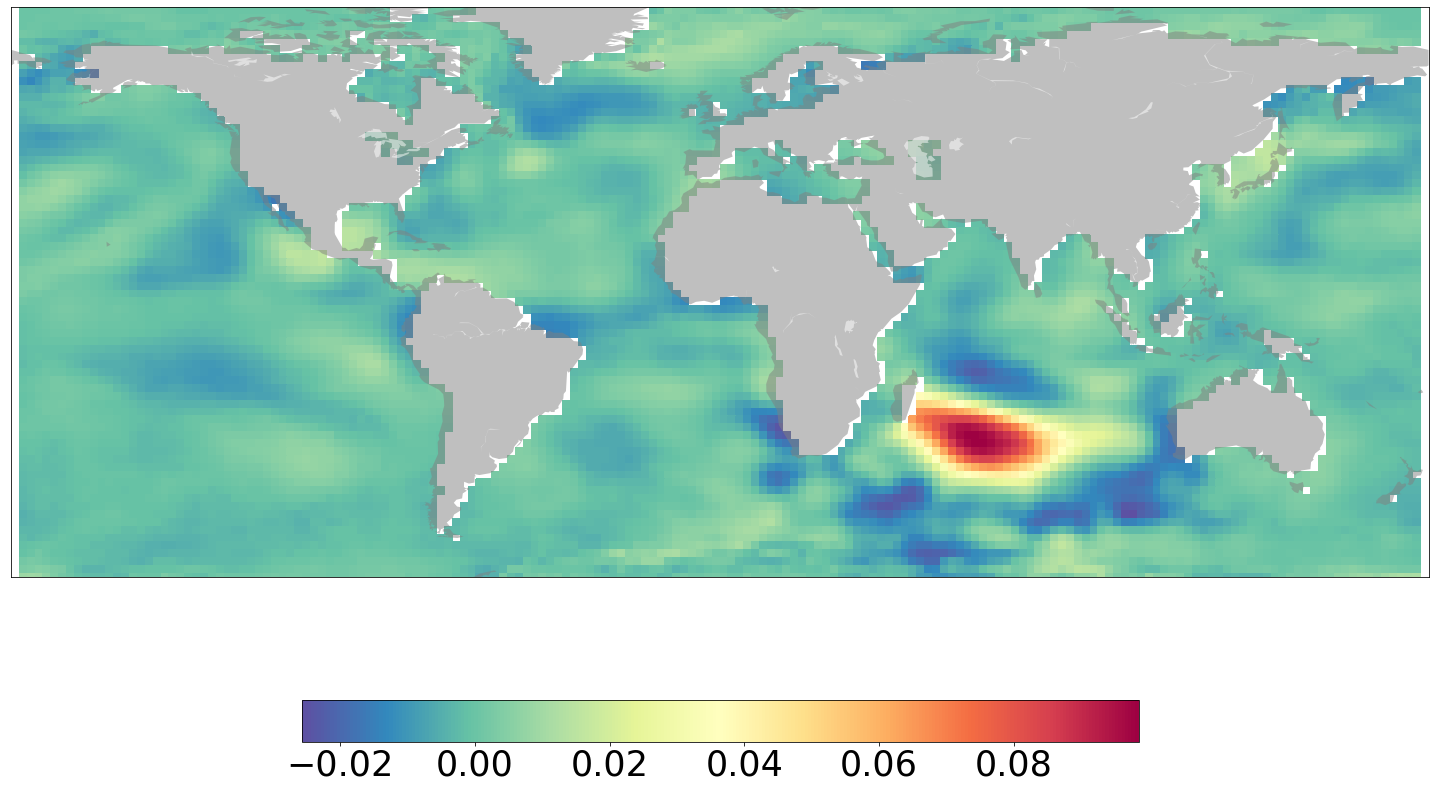}
\end{center}
\caption{The spatial structure of the 41st mode of variability of SST (EOF 41).}
\label{fig:EOF41}
\end{figure}

\begin{figure}
\begin{center}
\epsfxsize=1\hsize \epsfbox{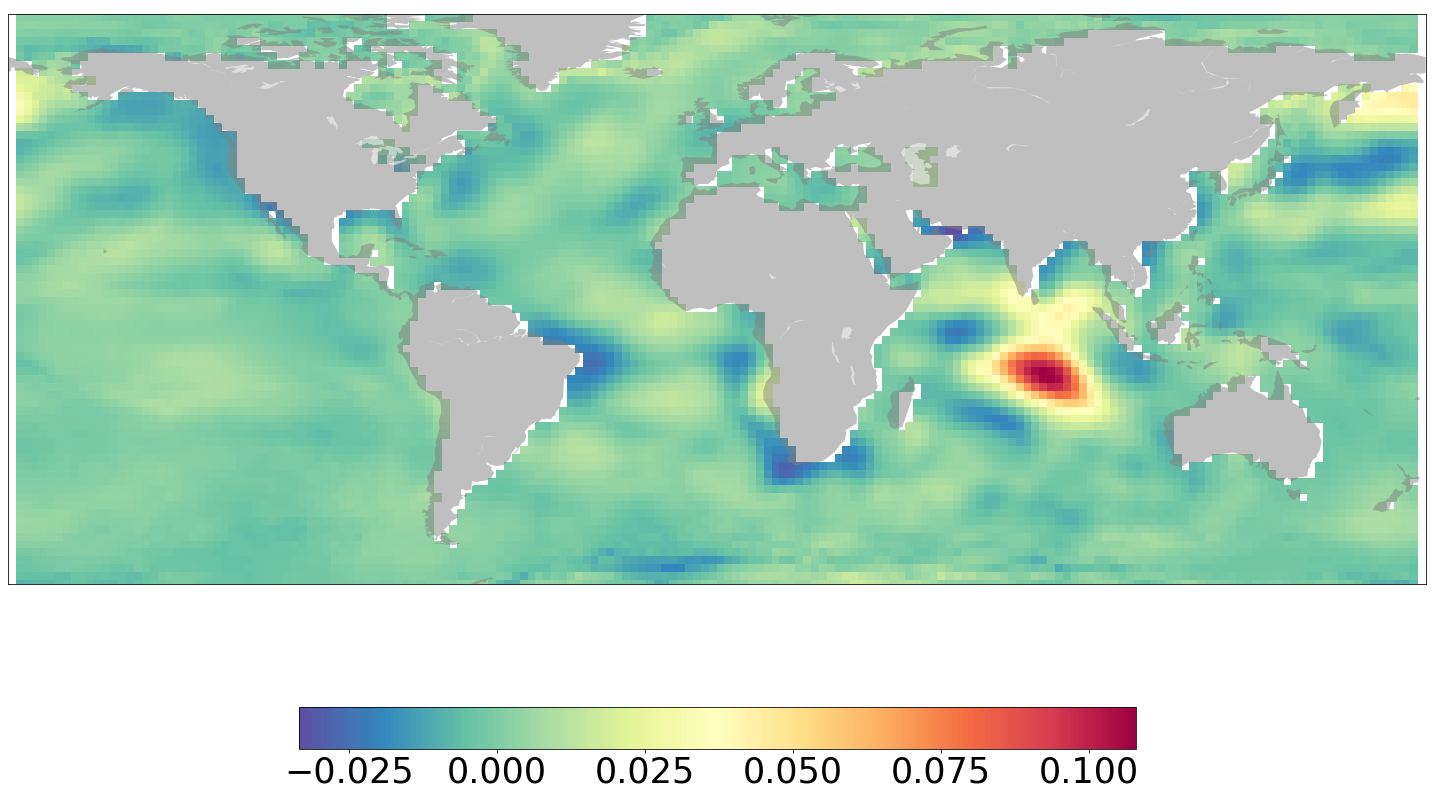}
\end{center}
\caption{The spatial structure of the 72nd mode of variability of SST (EOF 72).}
\label{fig:EOF72}
\end{figure}

The next causal links are the 41st and 72nd SST modes of variability. The former has a negative coefficient and lag of one month, and the latter has a positive coefficient with a lag of eleven months. As shown in Figures \ref{fig:EOF41} and \ref{fig:EOF72}, the EOFs of these two modes are in the Indian Ocean. We can draw a connection between these results and the past studies indicating the correlation between the SST anomaly in the Indian ocean and the deficiency of rainfall in the Eastern part of Africa.   

\begin{table}[]
\begin{tabular}{@{\extracolsep{10pt}}llll}
 Factor & Lag (Months)  & p value & Coefficient  \\[1ex] \hline 
 Drought & 1  & 0.00000  & +0.393  \\[1ex] 
 SST 2 & 12 & 0.00003 & -0.230 \\[1ex] 
 Drought & 12  & 0.00005  & -0.229 \\[1ex] 
 SST 41 & 1 & 0.00022  & -0.213 \\[1ex]
 SST 72 & 11 & 0.00261 & +0.176  
\end{tabular}
\caption{The influencing factors on drought in Ethiopia at $\alpha=0.05$ level.}
\label{table:results}
\end{table}

%SST 22 & 8 & 0.00233  & +0.175  \\[1ex]
% SST 52 & 10 & 0.00275 & +0.174   \\[1ex]
% SST 18 & 3 & 0.00359 & +0.168  \\[1ex]
% SST 58 & 4 & 0.00345 & +0.168  \\[1ex]
% SST 46 & 11 & 0.00725 & -0.156 

\subsection{Multiple Comparisons Effect}
Because we perform multiple simultaneous statistical tests, some of them may be significant due to sheer chance; therefore, we should keep the False Discovery Rate (FDR) at a fixed level by applying the Benjamini–Hochberg procedure \cite{benjamini1995controlling} which adjusts the p-values produced in MCI step. We compute the q-value as \cite{Runge2019InferringSciences}:
\begin{equation}
    q = min(p\frac{m}{r},1)
\end{equation}
where r is the rank of p-values when they are sorted in ascending order, and m is the total number of p-values. 
After applying the Benjamini–Hochberg procedure, the second PC of SST with a lag of 12 months and the q-value of 0.03781 is the only causal factor with statistically significant q-value ($<0.05$). In the next section, we cross-check this result by measuring this factor's predictive power as a feature in a drought forecasting model.

\subsection{Predictive Power of Second PC of SST}
We create a base model for Ethiopia drought prediction, using only the lagged copy of the drought time series as features. We quantify the predictive power of the second PC of SST based on the improvement to the performance after adding it as a feature to the base model.

To train the model, we use 360 months of data as the training set, which we denoted by $D_{train} = d_1, d_2, \dots d_T$ and 12 months as the test set denoted by $D_{test} = d_{T+1}, d_{T+2}, \dots d_{T+12}$ where $T=360$. This 372 months of data do not cover all available data because we want to apply cross-validation for time series. Like cross-validation for other data types, the final evaluation is the average values of several out-of-sample assessments on different proportions of the data set. 

We start by training the model on the first 360 months and evaluating it in the subsequent 12 months, and then we move forward by one month and repeat the same procedure in the new part of the dataset. We move forward until we reach the end of the dataset. This method is called rolling-window evaluation \cite{Tashman2000Out-of-sampleReview}. The final result is the average of 84 evaluations of the model on 84 subsets of the data.

We train the base model for one-step prediction by dividing the training part of the time series into several features/label observations such that at time step $t$ twelve time steps of $ d_{t-11},\dots,d_{t}$ are used to estimate $\hat{d}_{t+1}$.

To evaluate the model performance, we use a recurrent approach that uses the trained model to generate $\hat{d}_{T+1}$, and then take it as the ground truth, $\hat{d}_{T+1}\approx d_{T+1}$, and append it to $d_{T-10}$ to $d_{T}$ to estimate $\hat{d}_{T+2}$. We repeat this recurrent step until we obtain $\hat{d}_{T+12}$.    

To measure the predictive power of the second PC of SST which we denote it by $SST$, we simply add it as a feature to the model. Thus, in the training set, at any time step $t$, we append $SST_{t-11}$ to $ d_{t-11},\dots,d_{t}$ for estimating $\hat{d}_{t+1}$. In the testing step, because of the 12-months lag of $SST$, we have its true values from $T-11$ to $T$ to obtain $\hat{d}_{T+1},\dots,\hat{d}_{T+12}$.

We evaluate the model based on its ability to produce good estimations for $d_{T+1}\dots d_{T+12}$. We use the Root Mean Squared Error(RMSE), which is a metric used as "goodness-of-fit" measures in hydrologic and hydroclimatic model validation \cite{legates1999evaluating}, and it is defined as:

\begin{equation}\label{MSE}
RMSE(Y,\hat{Y})=\sqrt{\frac{1}{n} \sum_{i=1}^{n}(y_i - \hat{y}_i)^2}
\end{equation}
Where $n$ is the number of samples, $\hat{y}_i$ is the predicted value for data point i.

We use the following packages in Python: "Tensoreflow" \cite{tensorflow2015-whitepaper} for implementing neural network, "Scikit-learn" \cite{scikit-learn} for implementing Random Forest and Gradient Boosting and "CatBoost" \cite{dorogush2018catboost} for CatBoost implementation.   

Table \ref{table:ML} shows the result of the evaluations of the predictive model, with and without SST as a feature, using several machine learning algorithms. The performance improves when we add the second PC of SST to the models as an input variable. The best improvements are seen for the Neural Network and Random Forest models. The Gradient Boosting and CatBoost models show minimal improvement. The Neural Network model generates the minimum error among the models when it includes the second PC of SST as a predictor.

\begin{table}[]
\begin{tabular}{llll}
Model & RSME(without SST)  & RSME(with SST)\\ \hline 
Neural Network & 31.05 & \textbf{29.60} \\
Random Forest & 32.36 & \textbf{31.13} \\
Gradient Boosting & 36.45 & \textbf{36.31} \\
CatBoost & 31.62 & \textbf{31.34}  
\end{tabular}
\caption{Results of the evaluations of different models on datasets without/with the second PC of SST as a feature. Neural Network: two hidden layers each has eight nodes. Random Forest: max depth of 5 and n estimators of 100, Gradient Boosting: max depth of 4 and min samples split of 5, CatBoost: depth of 2}
\label{table:ML}
\end{table}

\subsection{Conclusion}
We find causal associations between the Global SST dataset and the drought time series in Ethiopia by applying the PCMCI algorithm. We discover a robust causal association from the PC of the second mode of variability of the SST to the number of grid points experiencing drought conditions in Ethiopia. The EOF of the second mode of variability has an enormous contribution from the ENSO region, and its coefficient is negative, which implies there is a higher probability that a large number of locations in Ethiopia experience drought conditions when the mean anomaly of SST in the ENSO region was at the cooling phase in the previous year. This result is consistent with the previous studies proving the same association for drought between 1999-2011 using correlation coefficient measures. We demonstrate that this association is likely to be causal, and it exists beyond the recent droughts in Ethiopia. 
Our study's most important finding is that there is a lag of 12 months involved in this causal association, which is helpful when trying to predict drought at socio-economically critical lead times (1-2 years). We find that the second PC of SST has predictive power, and it reduces the forecasting error when it is added as a feature to drought forecasting models.

Future studies can expand this analysis to other vulnerable regions of the world and include other influencing factors. These factors may be other meteorological anomalies, or they can be measures quantifying human interventions. Also, They can include the role of machine learning algorithms and why adding SST has a more significant impact on Neural Network or Random Forest.
\section*{Acknowledgments}
Funding for this project was provided by The Engineering and Physical Sciences Research Council (EPSRC) under the category of studentship with the reference number of 1943756.

\bibliographystyle{ieeetr}
\bibliography{ci_references,references}

\end{document}